\documentclass[12pt,epsf]{article}
\usepackage{amssymb,amsmath,amsbsy}
\usepackage{graphicx,color}
\pdfoutput=1

\newcommand{\be}{\begin{equation}}
\newcommand{\ee}{\end{equation}}
\newcommand{\bea}{\begin{eqnarray}}
\newcommand{\eea}{\end{eqnarray}}
\newcommand{\lb}{\left(}
\newcommand{\rb}{\right)}
\newcommand{\la}{\langle}
\newcommand{\ep}{\epsilon}
\newcommand{\ra}{\rangle}
\newcommand{\nn}{\nonumber}
\newcommand{\p}{\partial}
\newcommand{\mO}{{\mathcal O}}

\newcommand{\nbox}{{\,\lower0.9pt\vbox{\hrule \hbox{\vrule height 0.2 cm \hskip 0.19 cm \vrule height 0.2 cm}\hrule}\,}}

\def\href#1#2{#2}

\newcommand{\bi}{\begin{itemize}}
\newcommand{\ei}{\end{itemize}}
\newcommand{\ben}{\begin{enumerate}}
\newcommand{\een}{\end{enumerate}}
\newcommand{\bca}{\begin{cases}}
\newcommand{\eca}{\end{cases}}

\begin{document}
\begin{titlepage}
\vbox{
    \halign{#\hfil         \cr
           } 
      }  
\vspace*{15mm}
\begin{center}

{ \Large\bf 
Entanglement at a Scale and Renormalization Monotones}

\vspace*{15mm}
\vspace*{1mm}
Nima Lashkari
\vspace*{1cm}
\let\thefootnote\relax\footnote{$\mathrm{lashkari@mit.edu}$}

{{\small Center for Theoretical Physics, Massachusetts Institute of Technology\\
77 Massachusetts Avenue, Cambridge, MA 02139, USA \\
}}
%
%
\end{center}

\begin{abstract}

We study the information content of the reduced density matrix of a region in quantum field theory that cannot be recovered from its subregion density matrices. We reconstruct the density matrix from its subregions using two approaches: scaling maps and recovery maps. The vacuum of a scale-invariant field theory is the fixed point of both transformations.
We define the entanglement of scaling and the entanglement of recovery as measures of entanglement that are intrinsic to the continuum limit. Both measures increase monotonically under the renormalization group flow.  This provides a unifying information-theoretic structure underlying the different approaches to the renormalization monotones in various dimensions. Our analysis applies to non-relativistic quantum field theories as well the relativistic ones, however, in relativistic case, the entanglement of scaling can diverge.
\end{abstract}
\end{titlepage}

\vskip 1cm

\section{Introduction}

In recent years, the techniques and intuitions from quantum information-theory have proven to be immensely helpful in the study of many-body quantum systems. The entanglement structure of the low energy states of local Hamiltonians is a key concept in simulating lattice systems in condensed matter, the study of order parameters in phase-transitions, and constructing renormalization monotones in relativistic quantum field theories. 

The renormalization group (RG) flow is the process in which one integrates out the ultraviolet (UV) high energy degrees of freedom, and compensates for them by adjusting the coupling constants such that the low energy physics is unchanged. Since the information about the UV modes are washed out, one might expect that the RG flow is irreversible. RG monotones are functions that reflect this irreversability as they change monotonically under the flow.
 
The study of RG monotones in relativistic quantum field theory (QFT) was started by the seminal work of Zamolodchikov \cite{Zamolodchikov:1986gt}, where he showed that the two point function of stress tensor  in $2d$ QFT is a monotonic function of scale. In four dimensions, it was conjectured by Cardy in \cite{Cardy:1988cwa}, and later proved in \cite{Komargodski:2011vj}, that the $a$-anomaly term is an RG monotone. In two and three dimensions, the strong subadditivity (SSA) of entropy entropy was used to show that there are universal terms in the entanglement entropy of vacuum in QFT reduced to a ball-shaped region that are RG monotones \cite{Casini:2012ei}.
At the moment, the approaches to construct RG monotones seem to depend on the dimensionality of the spacetime, and a framework that works for all dimensions is missing.

In field theory, scaling is a unitary operation that allows us to compare the reduced density matrices on subsystems of different size. 
In this paper, we use scaling and the recovery maps of quantum information theory to quantify the amount of long-range quantum correlations at a scale. As a crucial step, we show that the Markov property of the vacuum of a conformal field theory implies that the vacuum state reduced to a null cone can be recovered perfectly from its subregions using both maps. 
We define the entanglement of scaling and the entanglement of recovery as two measures whose first derivative quantifies the {\it long-range entanglement}.\footnote{Intuitively, we think of the entanglement of scaling to be a generalization the measure introduced in \cite{Casini:2016udt} to general non-relativistic field theories.} Both of these functions increase monotonically under the RG flow. In some relativistic theories   the entanglement of scaling can be infinite; however, we expect that the entanglement of recovery to remain finite. Our monotonic functions are  generalizations of the $2d$ and $3d$ entanglement monotones to higher dimensions. They provide a unifying information-theoretic approach to RG monotones in various dimensions. Furthermore, it points to a connection between recovery maps in quantum information theory and the RG transformation of states that goes beyond the construction of monotones.\footnote{While this manuscript was in preparation, the papers \cite{Casini:2017roe,Casini:2017vbe} appeared, which have overlaps with some results presented here.} We start by reviewing some notions and tools in quantum information theory.


\subsection{Measuring asymmetry}
Consider a many-body finite quantum system split into  $n$ non-overlapping regions $A_1$ to $A_n$, with isomorphic Hilbert spaces on $A_i$.
The relabeling of the subsystem index $i$ is a unitary operation in the global Hilbert space: $\otimes_{i=1}^n\mathcal{H}_i$. A simple example of such a unitary is the translation defined by $i\to i+1 \mod n$:
\bea
&&U=\sum_{a_1\cdots a_n}|a_2\cdots a_n a_1\ra\la a_1\cdots a_n|\nn,
\eea
where $\{a_i\}$ is the basis that spans $\mathcal{H}_i$.
The density matrix $\rho_i$ on $A_i$ is mapped to $A_{i+1}$ with the local unitary 
\bea
&&\rho_{i+1}=\mathcal{E}(\rho_i)=U_i^\dagger \rho_i U_i\nn\\
&&U_i=\sum_{a_i,a_{i+1}}|a_{i+1}\ra \la a_i|.
\eea
If the transformation sends a subsystem $A$ to $\tilde{A}$, and the state is asymmetric under this transformation, some information about $\rho_A$ will be lost. 
The relative entropy $S(\rho_{\tilde{A}}\|\mathcal{E}(\rho_{A}))$ is a  measure of the amount of information in $\rho_A$ that is lost. 
It is non-negative, and vanishes if and only if $\rho_A$ is symmetric under the transformation.

\subsection{Measuring non-Markovianity}

Imagine that we are probing the global state with detectors that are localized in $A_1A_2$. The von Neumann entropy $S(\rho_{12})$ is a measure of the amount of quantum information $\rho_{12}$ is missing about a pure global state. If we made a larger detector that allows us access to the region $A_1A_2A_3$, then the new detector teaches us $S(A_3|A_1A_2)$ more qubits of information.  The quantity $S(A|A')\equiv S(A A')-S(A)$ is the conditional entropy. Another way to gain more information is by moving our detectors to adjacent sites $A_2 A_3$.  This gives us access to both $\rho_{12}$ and $\rho_{23}$; however, we are still missing the {\it long-range} correlations between $A_1$ and $A_3$. We would like to quantify the amount of quantum information (``entanglement") about in $\rho_{123}$ that is neither in $\rho_{12}$ nor in $\rho_{23}$. Naively, one can say that by moving the detector we have learned $S(A_3|A_2)$ but there are still  
\bea\label{SSA}
I(A_1:A_3|A_2)\equiv S(A_3|A_1A_2)-S(A_3|A_2)
\eea
more qubits in $\rho_{123}$ that we are missing. This quantity is the conditional mutual information (CMI), and is non-negative by the SSA inequality \cite{Lieb:1973cp}.

A careful study of the operational question of how well can one guess $\rho_{123}$ from the knowledge of $\rho_{12}$ and $\rho_{23}$ (the marginals) suggests that this naive estimate (CMI) is, indeed, a good measure of the amount of long-range entanglement. This can be seen from the two arguments below:

\begin{enumerate} 

\item Statistical physicist's prescription for the best guess is to consider the set of all consistent global states $\mathcal{C}$; that is all $\phi_{123}$ with $\phi_{12}=\rho_{12}$ and $\phi_{23}=\rho_{23}$. The best guess is a state $\phi_{123}$ in this set, which has the largest entropy \cite{Jaynes:1957zza}. 
It follows from the consistency condition that the entropy of the best guess is the CMI:
\bea
\sup_{\phi_{123}\in \mathcal{C}}S(\phi_{123})=I(A_1:A_3|A_2).
\eea

\item  Quantum information theorist's approach is to look at {\it recovery} maps. If a state has zero CMI, it can be reconstructed perfectly  from its marginals. Such states are called quantum Markov states, and satisfy the following property:
\bea\label{Markovident}
\log\phi_{123}=\log \phi_{12}+\log\phi_{23}-\log\phi_2.
\eea
The Markov state has no genuine long-range quantum correlations. All the correlations between $A_1$ and $A_3$ is classical and conditioned on $A_2$ \cite{HaydenPetz}. Furthermore, when the CMI is small one can use universal recovery maps to reconstruct the global state with high fidelity \cite{FawziRenner, WinterUniversal}. The CMI provides an upper bound on the fidelity distance of the recovered state. In fact, if we do not require the recovery map to be a quantum channel one can write down the  explicit map 
\bea
&&\rho_{recov}=e^{\log\rho_{12}+\log\rho_{23}-\log\rho_{2}}/Z,
\eea
that is hardly distinguishable from the global state:
\bea
&&S(\rho_{123}|\rho_{recov})\leq I(A_1:A_3|A_2).
\eea
Here $Z$ is the normalization of the state. 
The inequality above is satisfied trivially because $Z\leq 1$ \cite{Petz}.

\end{enumerate}

In our $n$-partite $A_1$ to $A_n$ example, if the state $\rho_{123}$ is Markovian one can recover it perfectly from $\rho_{12}$ and $\rho_{23}$, move the detector to the an adjacent site, and try to recover $\rho_{1234}$ from$\rho_{123}$ and $\rho_{34}$. This can be iterated to reconstruct $\rho_{1...m}$ for any $m<n$. If the state is recovered perfectly at each step, the global state is called a 
Quantum Markov chain \cite{Hastings,Czech:2014tva}. 
A quantum Markov chain found from adjacent local density matrices of size $r$ has the form
\bea
&&\log\rho_{1\cdots, m+r}=\log\rho_{m\cdots, m+r}+\nn\\
&& \sum_{k=1}^m(\log \rho_{k\cdots, k+r-1}-\log\rho_{k+1\cdots, k+r-1}).
\eea 
In our terminology, these Markov states have no entanglement at any scale larger than $r$.

Intuitively, a quantum Markov chain is scale-invariant, in the sense that all the information in a density matrix of size $R$ can be recovered perfectly from subsystems of size $r<R$. 
This suggests that quantum Markov states should appear naturally as the fixed points of the renormalization group flow.

\section{Entanglement of Scaling}
The states of a quantum field theory are wavefucntionals of fields: $\Psi(\phi(x))$. The transformations $f:x^\mu\mapsto x^\mu+\xi^\mu$ (diffeomorphisms) are the generalization of the relabeling operation in finite systems to the continuum limit. Analogously, diffeomorphisms act on the global state as unitary operators: $|\tilde{\psi}\ra=e^{i\int d\Sigma^\mu \xi^\nu T_{\mu\nu}}|\psi\ra$, where $\Sigma$ is the spacelike surface where the state lives, and $T_{\mu\nu}$ is the stress tensor. If we split the degrees of freedom into a subregion $A$ and the complement, then the unitary operator that maps the reduced state on $A$ to the reduced state to $\tilde{A}$ is:
\bea
U=\int [D\phi]_g |(f^{-1})^*\phi\ra \la \phi|
\eea
where $(f^{-1})^*$ is the pull-back of functions from $A$ to $\tilde{A}$ \cite{Faulkner:2016mzt}.

A familiar example of such diffeomorphisms is the generalization of translations in finite systems to the continuum limit. In quantum field theory, the translations are described by the unitaries $U=e^{i a^\mu P_\mu}$ which map $\rho_A$ to $\tilde{\rho}_{\tilde{A}}$:
\bea
\la\phi_a(x_\in A)|\rho_{A,g}|\phi_b(x\in A)\ra=\la (f^{-1})^*\phi_a|\rho_{\tilde{A},\tilde{g}} |(f^{-1})^*\phi_b\ra\nn,
\eea
where $\tilde{g}=(f^{-1})^*g$ is the transformed metric. If the translation is a symmetry of the background metric, and the state then the density matrix changes only by a unitary rotation.

\begin{figure}[b]
\centering
\includegraphics[width=0.59\textwidth]{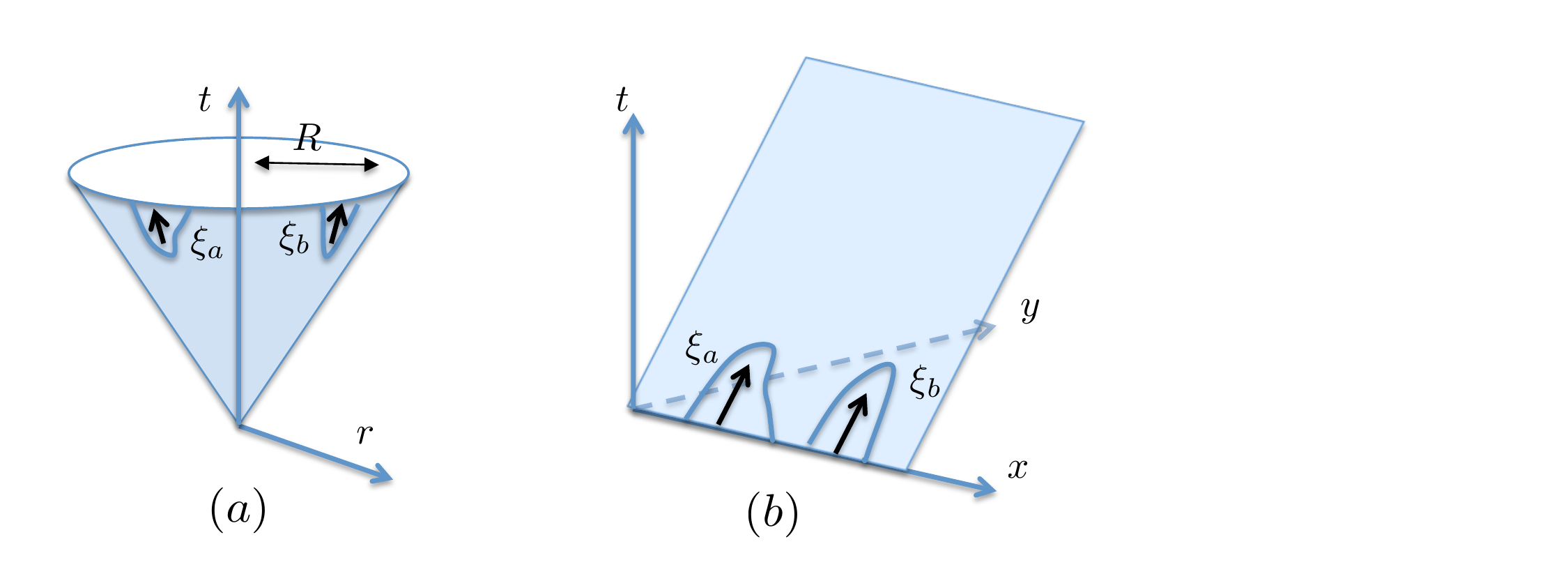}\\
\caption{\small{(a) Dilatataions that deform the boundary of ball at $t=R$, and act locally at particular angular variables $\Omega_a$ and $\Omega_b$ (b) Translations in the null direction that act locally in $x$ coordinates.}}
\label{fig1}
\end{figure}

In the remainder of this work, we will be interested in how local Dilatations acts on null cones. In polar coordinates, this maps $f:(t,r)\mapsto  (e^{\lambda(\Omega)} t,e^{\lambda(\Omega)} r)$, and leaves the perpendicular directions $\Omega$ untouched; see figure \ref{fig1}. Take a ball on the time slice $t=R$ centered at $r=0$. The boundary of this ball is on the null cone defined by $r-t=0$. The dilatation $f$ with constant $\lambda$ rescales the size of the ball from $R$ to $e^\lambda R$, and moves it from $t=R$ to $t=e^\lambda R$. The metric transforms by an overall conformal factor: $\tilde{g}=e^{2\lambda}g$. If the state is scale-invariant, for instance the vacuum of a scale-invariant theory, one can ignore the change of the metric, and the state remains unchanged up to a unitary.
To simplify the notation, we denote the unitarily scaled density matrix from $R$ to $R'$ by
\bea
\tilde{\rho}_{R'}\equiv\mathcal{E}(\rho_R)=U^\dagger \rho_R U,
\eea
where $R'$ has been suppressed in the notation, and will be clear from the context. 

We are interested in a quantum field theory that is a deformation of a scale-invariant theory by a relevant operator of scaling dimension $\Delta<d$
\bea
S_{QFT}=S_{scale-inv}+\lambda_0\int d^dx \mO(x),
\eea
where $\lambda_0=\mu^{\Delta-d}g_0$ is the dimensionful coupling at the UV length scale $\mu$. Diffeomorphism invariance allows us to compare $\rho_R$, the reduced states on a ball of size $R$, to a smaller ball $\rho_r$ rescaled back to $R$. In the UV ($r/\mu\ll 1$), the state $\rho_r$ can be approximated well by the scale-invariant vacuum state which transforms trivially under rescaling $\mathcal{E}$. 
In essence, the entanglement of scaling compares the reduced density matrix of a QFT to that of its ultraviolet fixed point.
 with corrections proportional to the coupling $\lambda_0$. 
 The modular operator of $\rho_r$ can be computed in the conformal perturbation theory.
It remains local in spacetime, to the first order in $\lambda_0$.
The relative entropy $S(\rho_R\|\mathcal{E}(\rho_r))$ is a measure of the amount of distinguishability lost under the dilatation. 
We define the {\it entanglement of scaling} to be
\bea
\mathcal{S}_{sc}(\rho_R)=\lim_{r\to 0}S(\rho_R\|\mathcal{E}(\rho_r)).
\eea
The entanglement of scaling is, by definition, non-negative. 
Similar to the entanglement entropy, the entanglement of scaling is invariant under any unitary operations: $\mathcal{S}_{sc}(\rho)=\mathcal{S}_{sc}(U^\dagger \rho U)$.

In essence, the relative entropy above compares the reduced density matrix of quantum field theory with that of its fixed point which was proposed as a C-function in relativistic quantum field theories in \cite{Casini:2016udt}. As the authors of \cite{Casini:2016udt} have discussed, this measure can be divergent in relativistic QFT for deformations that are not relevant enough.

%

\section{Markov states in QFT}


Take a quantum field theory density matrix $\rho_R$. If it is a quantum Markov state\footnote{In the remainder of this paper, we use the words Markov chain and Markov states synonymously.}, it can be perfectly recovered from its smaller marginals $\rho_r$, for any $r<R$. This suggests that there is no new physics at any length scale in between the $r$ and $R$. In other words, it is scale-invariant in that range. One might expect that the CFT vacuum reduced to ball-shaped regions are quantum Markov states. In this section, we show that this intuition is indeed correct.

%

Start with a ball-shaped region $A$ in a CFT vacuum state, and make two geometric deformations $f_a$ and $f_b$. The state will be Markovian if the CMI $I(\delta A_a,\delta_b A|A)$ vanishes for any finite size deformation. This quantity was computed in a perturbation theory in small deformations by \cite{Faulkner:2015csl}. 
They find the CMI to be 
\bea
I(\delta A_a;\delta A_b|A)=\delta A_a^{\bar{i}}\delta A_b^{(j)}\frac{2\pi^2 C_T}{(d+1)R^2}\frac{\eta_{\bar{ij}}}{|\Omega_a-\Omega_b|^{2(d-1)}},
\eea
where $\eta_{\bar{ij}}$ and $\delta A_a^{(i)}$ and $\delta A_b^{(j)}$ are, respectively, the metric and the area elements in the $t,r$ directions, and $C_T$ is the coefficient in the two-point function of the stress tensor. 
For a generic deformation, this CMI is non-zero. However, if we take the deformed ball to be on a null cone, that is $\xi=\xi^u(\Omega)\p_u$, the CMI is proportional to $\eta_{uu}$ which is zero in flat space. This leaves the possibility that for null deformations the vacuum state is Markovian. This was recently proved to be case in \cite{Casini:2017roe}. Here, we explore the Markov property from an intuitive tensor network point of view using the method of the Euclidean path-integrals.
In fact, it is pedagogical to start with a simpler example:\\

{\bf Ex. 1: QFT vacuum on half-space:} 
\vspace{.2cm}

As the first example, we show that the QFT vacuum in flat space reduced to a half-space is a quantum Markov state with respect to null deformations; see figure \ref{fig1}.
Consider the vacuum of a $d>2$ dimensional QFT in flat space $ds^2=du dv+dx^2+dz_i dz^i$, with $u=y+t$ and $v=y-t$ the null directions. We reduce the state to the region $A$, the $y>0$ half-space. The modular operator of this region, $K_A\equiv -\log\rho_A$, is local \cite{Casini:2011kv}. On the null surface $v=0$, it has the form
\bea
&&K_A\equiv -\log\rho_A=\int dx K_x\nn\\
&&K_x=\int d^{d-3}z \int_0^\infty du \: u T_{uu}(x).
\eea

\begin{figure}[b]
\centering
\includegraphics[width=0.59\textwidth]{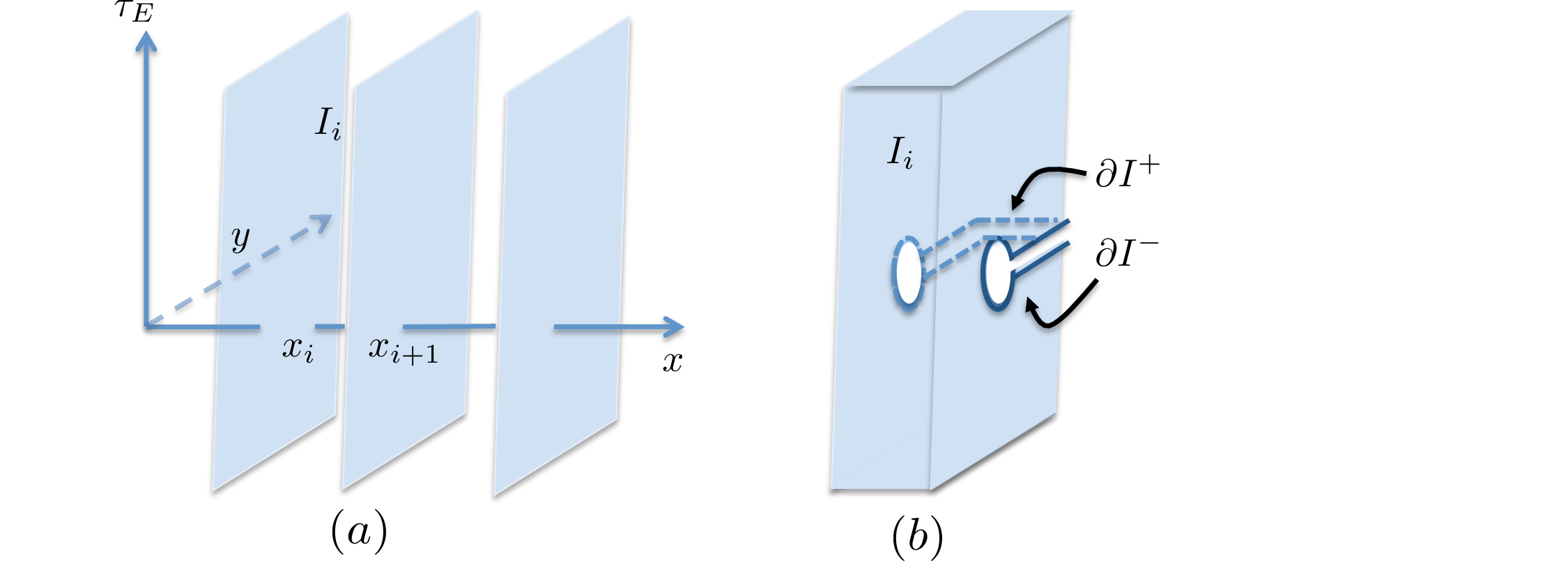}\\
\caption{\small{(a) Partitioning the Euclidean path-integral into slabs in the $x$ directions (b) The path-integral over each slab has five boundaries. Two boundaries at $x_i$ and $x_{i+1}$, two at $\p I^+$ and $\p I^-$ where the state lives, and one infinitesimal cylinder cut around the origin at $y=\tau_E=0$.}}
\label{fig2}
\end{figure}

In Euclidean QFT, the density matrix $\rho_A$ is represented by a path-integral on $\mathbb{R}^d$, with boundary conditions above and below $A$ in the Euclidean time; i.e. $(\tau_E=0^\pm,y>0)$ \cite{Calabrese:2004eu}.
One can split the $x$ direction into $n$ slabs $A_i=(x_i,x_{i+1})$, and insert the resolutions of identity in between slabs; see figure \ref{fig2}:
\bea\label{ref}
&&\rho=\int \prod_{i=1}^N [D\phi_i] \:\rho_{i}(\phi_i,\phi_{i+1}),\nn\\
&&\rho_{i}(\phi_i,\phi_{i+1})=\la\phi_i| \rho_i|\phi_{i+1}\ra.
\eea
Here, $\rho_{i}(\phi_i,\phi_{i+1})$ is an operator (transfer matrix) that acts only on the subsystem $A_i$. Intuitively, one can think of the expression in (\ref{matrixmult}) as a matrix product operator in the $x$ direction; see figure \ref{figfinal2}.

We apply a diffeomorphism that is non-zero only at $A_a$ and $A_b$, and deforms $A$ to $\tilde{A}=A+\delta_aA+\delta_b A$. The density matrix of $\tilde{A}$ is given by $\rho_{\tilde{A},\eta}= U^\dagger\rho_{A,g}U$,
 where $g_{\mu\nu}=\p_\mu \xi_\nu+\p_\nu\xi_\mu+\p_\mu \xi_\alpha\p_\nu \xi^\alpha$, and $\eta$ is the flat metric \cite{Faulkner:2016mzt}.
We take $f$ to be a translation in a null direction localized on two slabs $I_a$ and $I_b$: 
\bea
f_a:u\mapsto u +\lambda f(x_a),
\eea
with $f(x_a)$ a function that has a peak at the center of $A_a$, and goes to zero on the boundaries of $I_a$ at $x_a$ and $x_{a+1}$.\footnote{One might worry about the fact that the function $f$ is not infinitely differentiable. We will be ignorant of such subtleties here.}  The flat metric changes by $g_{x v}=\p_x\xi_v=\lambda\p_x f(x_a)$, which is nonzero only inside the slab $I_a$ and vanishes on the boundaries $\p I_a$. Partitioning the path-integral of $\tilde{\rho}_{\tilde{A}}$ according to (\ref{ref}) and comparing with $\rho_A$, only the transfer matrices $\rho_a$ and $\rho_b$ have changed. Let us focus on the matrix elements of one of these operators, $\tilde{\rho}_a$:
%
\bea
&&\la \phi^1(\p I_a^-)|\tilde{\rho}_a(\phi_a,\phi_{a+1})|\phi^2(\p I_a^+)\ra\nn\\
&&=\int_{\phi(x_a)=\phi_a, \phi(\p I_a^-)=\phi^1}^{\phi(x_{a+1})=\phi_{a+1},\phi(\p I_a^+)=\phi^2} [D\phi] e^{-S[\phi,g]},
\eea
where $\p I_a^\pm$ are the boundaries at $x\in A_a$ and $\tau_E=0^\pm$; see figure \ref{fig2}. The path-integral above is on $I_a$ that has five boundaries in the Euclidean $\mathbb{R}^{d+1}$. Two boundaries at $x=x_a$, $x=x_{a+1}$, two boundaries at $\p I_a^+$ and $\p I_a^-$, and a fifth boundary at $y^2+\tau_E^2=\ep$ which is a small cylinder cut around $y=\tau_E=0$. 

The only difference between the path-integrals for $\tilde{\rho}_a$ and $\rho_a$ is in the metric that goes into the action.
We Taylor expand the action around the flat space
\bea\label{rhok}
&&S[\phi,g]=\exp\lb \int_{I_a} \p^\mu \xi^\nu \frac{\delta}{\delta g^{\mu\nu}}\rb S[\phi,\eta]\nn\\
&&=\exp\lb -\int_{I_a}  \xi^\nu\p_\mu\frac{\delta}{\delta g^{\mu\nu}}+\int_{\p I_a} d\Sigma^\mu \xi^\nu\frac{\delta}{\delta g^{\mu\nu}}\rb S[\phi,\eta],\nn
\eea
where we have used the integration by parts, and $d\Sigma^\mu$ is the normal to the boundary $\p I_a$.
The term with the integral over $I_a$ vanishes, due to the fact that $\p_\nu \frac{\delta}{\delta g^{\nu\mu}}S[\phi,g]=\p_\nu T^{\mu\nu}$, which is identically zero. 

The change in the metric under the diffemorphism by $f_a$ is in the $g^{u x}$ component, and since $\xi^\mu$ has only $u$ components, only the two boundaries at constant $x$ contribute to (\ref{rhok}). However, we chose $\xi$ to vanish on these boundaries; therefore $S[\phi,g]$ on $I_a$ can be replaced with its flat space value $S[\phi,\eta]$.
Hence, the transfer matrices in the partitioned path-integral in (\ref{ref}) do not change:
\bea
\tilde{\rho}_{a}(\phi_a,\phi_{a+1})=\rho_a(\phi_a,\phi_{a+1}).
\eea
Hence, there is a unitary that rotates the overall density matrix $\rho_A$ to $\tilde{\rho}_{\tilde{A}}$:
\bea\label{matrixmult}
\tilde{\rho}_{\tilde{A}}=(\mathbb{I}\otimes U_a^\dagger\otimes U^\dagger_b)\rho_{A}(\mathbb{I}\otimes U_a\otimes U_b)\ .
\eea
This unitary operator is $U_a(x)=e^{i \alpha Q_a}$ where $Q_a=\int du\: T_{uu}(a)$ is the average null energy operator. 

\begin{figure}
\centering
\includegraphics[width=.9\textwidth]{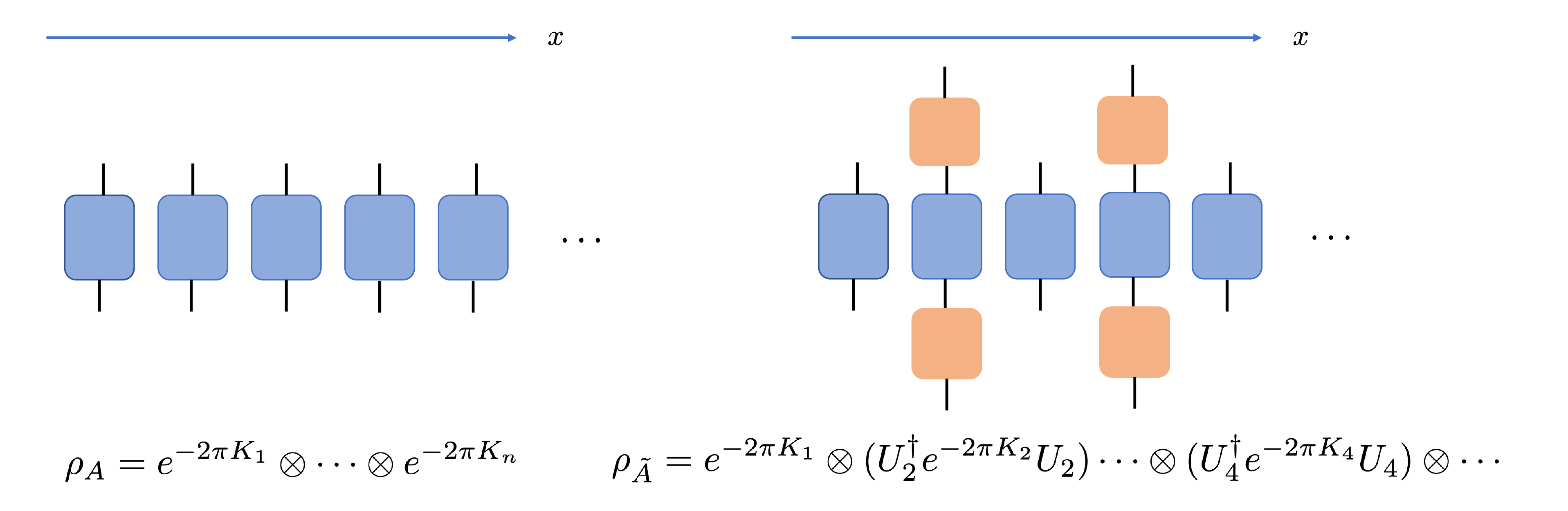}\\
\caption{\small{(a) The density matrix of the half-space on a null sheet factorizes in free field theory (b) a shape deformation on the null sheet at point $x=a$ corresponds to acting with unitaries $U_a$.}}
\label{figfinal}
\end{figure}

In the null quantization of free field theory, the vacuum state is the zero eigenvector of the null momentum $P_u$. Furthermore, we know that this state is a tensor product of the vacuua of the $Q_x$:
\bea
&&|\Omega\ra=\otimes_x |\Omega_x\ra,\qquad Q_x|\Omega_x\ra=0\ .
\eea
This means that the reduced density matrix of half-space is also a tensor product
\bea
&&\rho=\otimes_x \rho_x=\otimes_x e^{-2\pi K_x}
\eea
where $\rho_x$ is the vacuum density matrix on the half-space found from the ground state $|\Omega_x\ra$. There is no entanglement between $\rho_x$ and $\rho_{x'}$ and the matrix product operator is of the form in figure \ref{figfinal}. It is clear that applying the unitaries $U_a$ and $U_b$ only changes the matrices $\rho_a$ and $\rho_b$ and cannot create entanglement. Therefore, it is trivially true in free theory that
\bea
K_{\tilde{A}}=K+(U_a^\dagger K_a U_a-K_a)+(U_b^\dagger K_b U_b-K_b).\nn
\eea
The two-dimensional Poincare group gives us the commutation relation
\bea
[K_x,Q_a]=-i Q_a \delta(x-a)\ .
\eea
which results in a resummation of the Baker-Campbell-Hausdorff expansion:
\bea
U_x^\dagger e^{-2\pi K_x} U_x= e^{-2\pi (K_x-\alpha Q_x)}\ .
\eea
As a result, the modular Hamiltonian of the deformed region is
\bea\label{MarkovMod}
K_{\tilde{A}}=K_A-\alpha (Q_a-Q_b)\ .
\eea
This is the Markov property of vacuum in free field theory as was originally argued for in \cite{Wall:2011hj}.
\begin{figure}
\centering
\includegraphics[width=.9\textwidth]{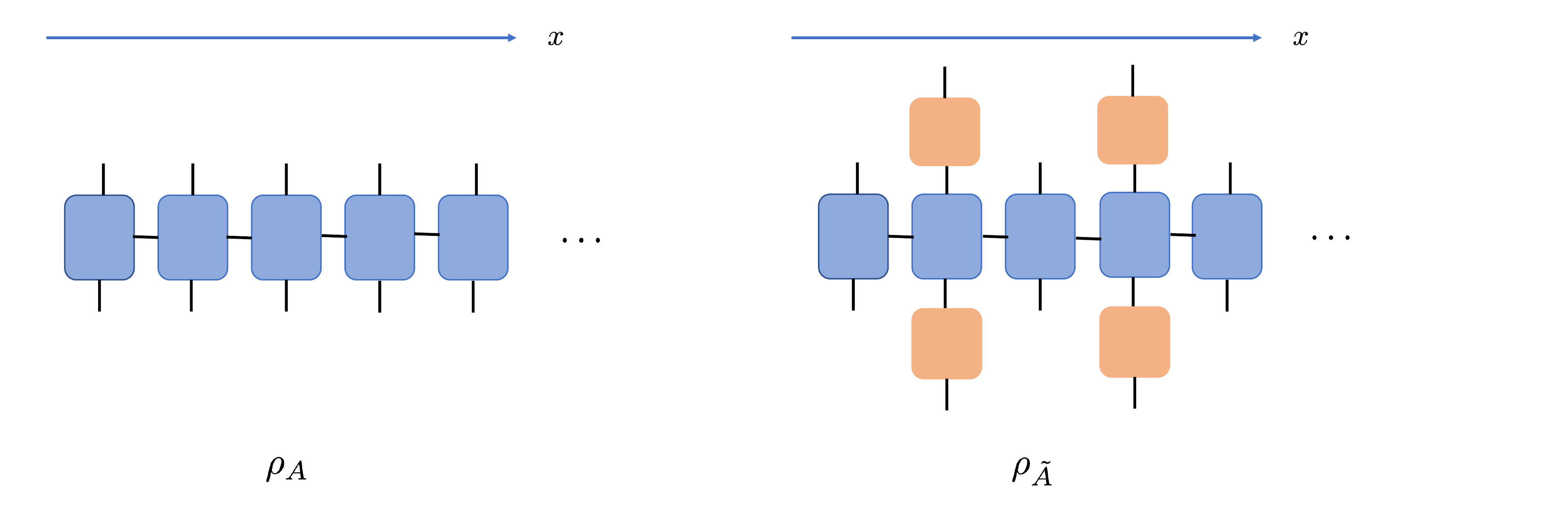}\\
\caption{\small{(a) The density matrix of the half-space on a null sheet factorizes in interacting theories is entangled in the $x$ direction (b) a shape deformation on the null sheet at point $x=a$ corresponds to acting with unitaries $U_a$.}}
\label{figfinal2}
\end{figure}

In a general interacting theory the vacuum state is the zero eigenvector of $Q_x$ smoothed in the $x$ direction.
However, we expect $Q_x$ with no smoothing to have no normalizable zero eigenvector.\footnote{We thank Juan Maldacena for pointing this out to us.} This is reflected in the fact that the vacuum state is entangled across cuts of constant $x$. The matrix product operator representation of the vacuum density matrix is schematically drawn in figure \ref{figfinal2}. 
The density matrix is still
\bea
\rho_A= e^{-2\pi K_1}e^{-2\pi K_2}\cdots e^{-2\pi K_n}
\eea
which is not a product state.
It has been argued in \cite{Casini:2017roe} that the commutator 
\bea
[K_x,Q_a]=-i Q_a \delta(x-a)\ .
\eea
remains unmodified in interacting theories. 
One can commute the operators $e^{i \alpha Q_x}$ with $e^{-2\pi K_{x'}}$ and finds the same expression for the modular Hamiltonian as in the free theory:
\bea
K_{\tilde{A}}=K_A-\alpha (Q_a-Q_b),
\eea
which is the Markov property of the vacuum density matrix on a null sheet.\\

{\bf Ex. 2: CFT vacuum on a null cone:} 
\vspace{.2cm}

There is a conformal transformation that maps the causal development of a half-space $A$ to the causal development of a ball $B$ \cite{Casini:2011kv}. If $K_A$ and $K_B$ are, respectively, the modular operators of subsystems $A$ and $B$, there exists a unitary such that $K_B=U^\dagger K_A U$. Under this conformal transformation, the deformed half-space $A+\delta_a A$ is mapped to a deformed ball $B+\delta_a B$; see figure \ref{fig1}. Deformations on the null surface in $A$ are sent to deformations of $B$ on the null-cone. The equation (\ref{MarkovMod}) with $\tilde{A}$ continues to hold for the vacuum of a CFT in arbitrary dimensions with $\tilde{A}$ a deformation of the ball on the null cone that is its causal development.
As a result, the vacuum of a $d$-dimensional CFT is a quantum Markov state with respect to deformations on a null cone.

In 2d CFTs, any state that is a descendant of vacuum with arbitrary time-dependence is related to vacuum by a conformal transformation, and remains a quantum Markov state. It is straightforward to check that SSA is saturated in these states from the expressions in \cite{Roberts:2012aq}.\footnote{We thank Matthew Roberts for pointing this out to us.}


\subsection*{Near Markov States}
Before applying the SSA inequality to the states of a quantum field theory, we would like to have an analogue of CMI that is insensitive to the ultraviolet details. We replace the entanglement entropies in CMI with the entanglement of scaling:
\bea
&&I_{sc}(A_1:A_3|A_2)\equiv\nn\\
&&\mathcal{S}_{sc}(\rho_{12})+\mathcal{S}_{sc}(\rho_{23})-\mathcal{S}_{sc}(\rho_{2})-\mathcal{S}_{sc}(\rho_{123})\nn\\
&&=I_{\rho_R}(A_1:A_3|A_2)-\lim_{r\to 0}I_{\rho_r}(A_1:A_3|A_2)\nn\\
&&=I(A_1:A_3|A_2)\geq 0,
\eea
where we have used the fact that the UV CFT state is Markovian. Note that in relativistic quantum field theory there is no guarantee that this quantity remains finite term by term. 

\section{Entanglement at a Scale}


In this section, for simplicity we restrict to vacuum state of QFTs in flat space.\footnote{The generalization of the measures introduced here to arbitrary states requires minor, but straightforward modifications.} The goal is to find an information-theoretic measure that quantifies the entanglement at a scale that is insensitive to the UV and has an operational interpretation. A measure of entanglement at scale $R$ is a function that $\rho_R$ and its derivatives $\p_R^m\rho_R$. Here, we compare three candidate measures that appear natural from an information-theory point of view:  

\begin{enumerate}

\item The obvious candidate is the relative entropy $S(\rho_{R+\delta R}|\mathcal{E}(\rho_{R}))$. This quantity vanishes at the first order in $\delta R$, due to the smoothness of relative entropy. At the second order, it becomes the quantum Fisher information which is a metric in the space of density matrices:
\bea
S(\rho_{R+\delta R}|\rho_R)=(\delta R)^2 \la \delta^R\rho,\delta^R\rho\ra_R+O((\delta R)^3).\nn
\eea 
It is finite, non-negative at any $R$, and vanishes in CFTs. It is a metric, and hence satisfies the triangle inequality. Quantum Fisher information has an interpretation in terms of distinguishability, as it is the variation of a relative entropy. 

\item The second candidate is  the derivative $\p_R \mathcal{S}_{sc}(\rho_R)$. It is finite, and non-negative at any $R$ (see the supplementary material for a proof):
\bea
\p_R \mathcal{S}_{sc}(R)\geq 0.
\eea
This quantity is expected to be insensitive to the UV details, and has the benefit that its integral, $\mathcal{S}_{sc}$, resembles a  smoothed-out version of $S_{UV}-S_{IR}$.  However, in relativistic field theory it diverges for deformations that are not relevant enough.

\item The third candidate, the information-theorist's favorite, is based on recovery maps and SSA. The task is to quantify how well one can recover the state $\rho_{R+\delta R}$ from the knowledge of all balls of size $R$ within the causal development of $\rho_{R+\delta R}$. That is to say, we want to build a ball of size $R+\delta R$ from the iteration of a recovery map which acts on balls of size $R$. One way to do this was introduced in  \cite{Casini:2012ei}. 
Take two balls with boundaries on a null cone. As we bring the balls close in the angular directions on the cone,the distance between $\delta_aA$ and $\delta_b A$ tends to $R$. the CMI measures the entanglement at scale $R$. To obtain the larger $\rho_{R+\delta R}$ we have to apply the recovery map many times following \cite{Casini:2012ei}, and add up the CMI contributions at each step.
The total sum of the CMI we obtain as we repeat this recipe is the quantity that we define to be the derivative of the {\it entanglement of recovery}
\bea
\p_R\mathcal{S}_{rec}(\rho_R)\equiv\lb (d-3)\p_R+R\p_R^2\rb \mathcal{S}_{sc}(R)\geq 0.
\eea
It is a measure of the entanglement in the vacuum of QFTs at the scale $R$, that has an operational interpretation in terms of recovery. It vanishes in a CFT vacuum. Integrating this quantity from the UV to the scale $R$ we obtain 
\bea
&&\mathcal{S}_{rec}(R)=(d-2-R\p_R)\mathcal{S}_{sc}(\rho_R).
\eea
\end{enumerate}

\section{Renormalization monotones}

We are encouraged by \cite{Casini:2015woa} to look for an RG monotone in arbitrary dimensions that has the following properties
\begin{enumerate}
\item It is a finite dimensionless quantity, and regularization independent.

\item It decreases monotonically along the flow.  

\item If the flow ends in an IR fixed point, the value of the function can only depend on quantities that are intrinsic to the UV and IR fixed points.
\end{enumerate}

We expect both the entanglement of scaling and the entanglement of recovery to satisfy the first property in non-relativistic examples. In relativistic theories, the conditions under which they remain finite is unclear to us and deserves further study.
Both measures satisfy the second criterion:
\bea
&&\p_R S_{sc}(R)\geq 0\nn\\
&&\p_R S_{rec}(R)\geq 0.
\eea
In all the known examples in $2d$ and $3d$ they also satisfy the third criterion. It is unclear to us, whether this continues to be the case in all dimensions.  

In $2d$ and $3d$ they do indeed reduce to all the known monotones. 
The entanglement of scaling, $\mathcal{S}_{sc}(R)$, is a smoothed version of the RG monotone defined in \cite{Casini:2016udt}, which is the relative entropy of vacuua in two different CFTs. While intuitive, the smoothness of $\mathcal{S}_{sc}(R)$ deserves further investigation. We believe that studying the entanglement of scaling in more detail can shed light on the UV divergences in the quantity in \cite{Casini:2016udt} for the particular range of the deformation scaling dimensions $\Delta>(d+2)/2$.

The entanglement of recovery, $\mathcal{S}_{rec}(R)$, is a smoothed version of the entanglement monotones in 2d and 3d introduced in \cite{Casini:2012ei} generalized to arbitrary dimension. As this work was in its final stages, we learned about the work in \cite{Casini:2017vbe} that generalizes the previous entanglement proof to the a-theorem in four dimensions. It is of great interest to relate the entanglement of recovery to other known quantities of CFTs in $d>4$.

\section{Conclusions}
In this work, we studied a connection between recovery maps in quantum information theory, and the renormalization group flow in quantum field theories. Applying information-theoretic tools, and taking advantage of the diffeomorphism invariance of QFT, we constructed candidate functions for the entanglement at a scale. Two new entanglement measures intrinsic to the continuum limit,  the entanglement of scaling and the entanglement of recovery were defined. They are built such that their first derivatives in scale quantifies the amount of entanglement at scale. However, the more natural quantity from the point of view of the recovery maps is the entanglement of recovery. Both quantities are monotonic under a change of scale. A better understanding of the RG monotones in higher dimensions can be achieved by studying these quantities and relating them to the properties of the IR scale-invariant fixed point. 

It is tempting to rewrite the entanglement of scaling in the language of the algebraic QFT as
\bea
\lim_{\lambda\to 0}\la \Omega|\Delta_{\Omega,U_\lambda^\dagger \Omega U_\lambda}|\Omega\ra,
\eea
and avoid referring to the density matrix. Here, $|\Omega\ra$ is the state of a QFT, and $\Delta_{\Omega,\Omega'}$ is the relative modular operator of the two states with respect to a region, and $U_\lambda$ generates dilatation by factor $\lambda$. We postpone a further investigation of this, and potential connections between the entanglement of scaling and the renormalized entanglement entropy \cite{Liu:2012eea} to future work.
Furthermore, since our approach views RG as an operation on a QFT state, the RG monotones we find characterize a particular flow from the UV to the IR. An interesting question to explore is whether this quantity can be read off, directly from a CFT Hilbert space. 
%

\section{Acknowledgements}

We are greatly indebted to Hong Liu for many valuable discussions on renormalization group flow. Also, we would like to thank Laurent Chaurette, Matthew Headrick, Petr Kravchuk, Juan Maldacena, Srivatsan Rajagopal and Matthew Roberts for informative discussions.

\appendix

\section{The entanglement of scaling is monotonic}\label{AppA}
We are interested in the derivative:
\bea
\lim_{\mu\to 0}\p_R S(\rho_R\|\mathcal{E}(\rho_\mu))\geq 0.
\eea

We start by proving that the operations, $\mathcal{E}$ and $\mathcal{N}$ commute: $\mathcal{N}(\mathcal{E}(\rho))=\mathcal{E}(\mathcal{N}(\rho))$. Split the system in two parts: the part that is traced out $A$, and the remaining part $B$. The matrix elements of $\mathcal{E}(tr_A\rho)$ are
\bea
&&\int [D\psi]_A \:\la\psi_A (f^{-1})^*\phi_B^+|\rho|\psi_A (f^{-1})^*\phi_B^-\ra.
\eea
After a change of variables this is equal to
\bea
\int [D(f^{-1})^*\psi]_A \:\la(f^{-1})^*\psi_A (f^{-1})^*\phi_B^+|\rho|(f^{-1})^*\psi_A (f^{-1})^*\phi_B^-\ra.\nn
\eea
which is nothing but $tr_A\mathcal{E}(\rho)$.

Relative entropy is monotonic under  a partial trace: $\mathcal{N}_{R\to R-\delta R}$. We have
\bea
&&S\lb \rho_R\|\mathcal{E}(\rho_\mu)\rb\geq S\lb \mathcal{N}(\rho_R)\|\mathcal{N}\mathcal{E}(\rho_\mu)\rb\nn\\
&&=S\lb\mathcal{N}(\rho_R)\|\mathcal{E}(\mathcal{N}(\rho_\mu))\rb\nn\\
&&=S(\rho_{R-\delta R}\|\mathcal{E}(\rho_\mu)+\mu\mathcal{E}(\delta\rho_\mu))
\eea
Taking the limit $\mu\to 0$ we establish that 
\bea
\p_R\mathcal{S}_{sc}(R)\geq 0.
\eea

\bibliographystyle{apsrev4-1}


\begin{thebibliography}{50}

\bibitem{Zamolodchikov:1986gt} 
  A.~B.~Zamolodchikov,
  ``Irreversibility of the Flux of the Renormalization Group in a 2D Field Theory,''
  JETP Lett.\  {\bf 43}, 730 (1986)
  [Pisma Zh.\ Eksp.\ Teor.\ Fiz.\  {\bf 43}, 565 (1986)].

\bibitem{Cardy:1988cwa} 
  J.~L.~Cardy,
  ``Is There a c Theorem in Four-Dimensions?,''
  Phys.\ Lett.\ B {\bf 215}, 749 (1988).
  doi:10.1016/0370-2693(88)90054-8

\bibitem{Komargodski:2011vj} 
  Z.~Komargodski and A.~Schwimmer,
  ``On Renormalization Group Flows in Four Dimensions,''
  JHEP {\bf 1112}, 099 (2011)
  doi:10.1007/JHEP12(2011)099
  [arXiv:1107.3987 [hep-th]].

\bibitem{Casini:2012ei} 
  H.~Casini and M.~Huerta,
  ``On the RG running of the entanglement entropy of a circle,''
  Phys.\ Rev.\ D {\bf 85}, 125016 (2012)
  doi:10.1103/PhysRevD.85.125016
  [arXiv:1202.5650 [hep-th]].

\bibitem{Casini:2017roe} 
  H.~Casini, E.~Teste and G.~Torroba,
  ``Modular Hamiltonians on the null plane and a Markov property of the vacuum state,''
  arXiv:1703.10656 [hep-th].

\bibitem{Casini:2017vbe} 
  H.~Casini, E.~Teste and G.~Torroba,
  ``The a-theorem and the Markov property of the CFT vacuum,''
  arXiv:1704.01870 [hep-th].
  
 
\bibitem{Lieb:1973cp} 
  E.~H.~Lieb and M.~B.~Ruskai,
  ``Proof of the strong subadditivity of quantum-mechanical entropy,''
  J.\ Math.\ Phys.\  {\bf 14}, 1938 (1973).
  doi:10.1063/1.1666274
  
\bibitem{Jaynes:1957zza} 
  E.~T.~Jaynes,
  ``Information Theory and Statistical Mechanics,''
  Phys.\ Rev.\  {\bf 106}, 620 (1957).
  doi:10.1103/PhysRev.106.620
  

\bibitem{HaydenPetz}
P.~Hayden, R.~Jozsa, D.~Petz, A.~Winter, 
``Structure of states which satisfy strong subadditivity of quantum entropy with equality"
Communications in mathematical physics, 246(2), pp.359-374.

\bibitem{FawziRenner}
O.~Fawzi, R.~ Renner, 
``Quantum conditional mutual information and approximate Markov chains." 
arXiv preprint arXiv:1410.0664 (2014)

\bibitem{WinterUniversal}
M.~Junge, R.~Renner, D.~Sutter, M.~Wilde, A.~Winter, 
``Universal recovery from a decrease of quantum relative entropy"
preprint arXiv:1509.07127.


\bibitem{Petz} 
  D.~Petz,
  Springer Science and Business Media, 2007
  
  \bibitem{Hastings}
  D.~Poulin, M.~B.~Hastings, 
  ``Markov entropy decomposition: a variational dual for quantum belief propagation"
  
  Physical review letters, 106(8), 080403.

\bibitem{Czech:2014tva} 
  B.~Czech, P.~Hayden, N.~Lashkari and B.~Swingle,
  ``The Information Theoretic Interpretation of the Length of a Curve,''
  JHEP {\bf 1506}, 157 (2015)
  doi:10.1007/JHEP06(2015)157, 10.1007/jhep06(2015)157
  [arXiv:1410.1540 [hep-th]].

\bibitem{Faulkner:2016mzt} 
  T.~Faulkner, R.~G.~Leigh, O.~Parrikar and H.~Wang,
  ``Modular Hamiltonians for Deformed Half-Spaces and the Averaged Null Energy Condition,''
  JHEP {\bf 1609}, 038 (2016)
  doi:10.1007/JHEP09(2016)038
  [arXiv:1605.08072 [hep-th]].


\bibitem{Faulkner:2015csl} 
  T.~Faulkner, R.~G.~Leigh and O.~Parrikar,
  ``Shape Dependence of Entanglement Entropy in Conformal Field Theories,''
  JHEP {\bf 1604}, 088 (2016)
  doi:10.1007/JHEP04(2016)088
  [arXiv:1511.05179 [hep-th]].

\bibitem{Casini:2011kv} 
  H.~Casini, M.~Huerta and R.~C.~Myers,
  ``Towards a derivation of holographic entanglement entropy,''
  JHEP {\bf 1105}, 036 (2011)
  doi:10.1007/JHEP05(2011)036
  [arXiv:1102.0440 [hep-th]].

\bibitem{Calabrese:2004eu} 
  P.~Calabrese and J.~L.~Cardy,
  ``Entanglement entropy and quantum field theory,''
  J.\ Stat.\ Mech.\  {\bf 0406}, P06002 (2004)
  doi:10.1088/1742-5468/2004/06/P06002
  [hep-th/0405152].

\bibitem{Wall:2011hj} 
  A.~C.~Wall,
  ``A proof of the generalized second law for rapidly changing fields and arbitrary horizon slices,''
  Phys.\ Rev.\ D {\bf 85}, 104049 (2012)
  Erratum: [Phys.\ Rev.\ D {\bf 87}, no. 6, 069904 (2013)]
  doi:10.1103/PhysRevD.87.069904, 10.1103/PhysRevD.85.104049
  [arXiv:1105.3445 [gr-qc]].

\bibitem{Roberts:2012aq} 
  M.~M.~Roberts,
  ``Time evolution of entanglement entropy from a pulse,''
  JHEP {\bf 1212}, 027 (2012)
  doi:10.1007/JHEP12(2012)027
  [arXiv:1204.1982 [hep-th]].
  
\bibitem{Casini:2015woa} 
  H.~Casini, M.~Huerta, R.~C.~Myers and A.~Yale,
  ``Mutual information and the F-theorem,''
  JHEP {\bf 1510}, 003 (2015)
  doi:10.1007/JHEP10(2015)003
  [arXiv:1506.06195 [hep-th]].
  
\bibitem{Casini:2016udt} 
  H.~Casini, E.~Teste and G.~Torroba,
  ``Relative entropy and the RG flow,''
  JHEP {\bf 1703}, 089 (2017)
  doi:10.1007/JHEP03(2017)089
  [arXiv:1611.00016 [hep-th]].
  
\bibitem{Liu:2012eea} 
  H.~Liu and M.~Mezei,
  ``A Refinement of entanglement entropy and the number of degrees of freedom,''
  JHEP {\bf 1304}, 162 (2013)
  doi:10.1007/JHEP04(2013)162
  [arXiv:1202.2070 [hep-th]].
  
\end{thebibliography}

\end{document}